%% file: pb_updfevolv_note.tex
\documentclass[11pt]{article} 
\usepackage{mystyle-new}
\usepackage{authblk}
\usepackage[normalem]{ulem}
\usepackage[T1]{fontenc}
\usepackage{epsfig,amsmath} 
\usepackage{hyperref}
\usepackage{color}
\usepackage{graphicx}
\usepackage{orcidlink}
\usepackage{cancel}
\usepackage{cprotect}

\definecolor{red}{rgb}{1,0,0}
\def\lesssim{\ \hbox{\raise 2pt \hbox{$<$} \kern -13pt
                     \lower 3pt \hbox{$\sim$}}\ }
\def\greatersim{\ \hbox{\raise 2pt \hbox{$>$} \kern -13pt
                     \lower 3pt \hbox{$\sim$}}\ }

\def\lsim{\mathrel{\rlap{\lower4pt\hbox{\hskip1pt$\sim$}}
    \raise1pt\hbox{$<$}}}                
\def\gsim{\mathrel{\rlap{\lower4pt\hbox{\hskip1pt$\sim$}}
    \raise1pt\hbox{$>$}}}                

\newcommand{\dd}{\mathrm{d}}

\input epsf.tex
\def\desepsf(#1 width #2){\epsfxsize=#2 \epsfbox{#1}}

\def\qt{\ensuremath{q_{\rm t}}}

\def\zM{\ensuremath{z_{ M}}}

\newcommand{\alphas}{\ensuremath{\alpha_\mathrm{s}}}

\newcommand{\PZ}{{\rm Z}}

\newcommand{\asmz}{\ensuremath{\alphas(m_{\PZ})}}

\newcommand{\PB}{PB}

\newcommand{\qcdnum}{{\sc QCDnum}}
\newcommand{\apfel}{{\sc Apfel}}

  \newenvironment{defl}[1]%
  {\begin{list}{}{\settowidth{\labelwidth}{#1}%
  \setlength{\leftmargin}{\labelwidth}%
  \addtolength{\leftmargin}{\labelsep}%
  \setlength{\itemsep}{0pt plus 1pt}
  \setlength{\parsep}{0pt plus 1pt}
  \setlength{\topsep}{0pt plus 1pt}
  \setlength{\partopsep}{0pt plus 1pt}
  \setlength{\parskip}{2mm plus 1mm minus 1mm}
  }}%
  {\end{list}}

\newenvironment{tolerant}[1]{\par\tolerance=#1\relax}{ \par }
\usepackage{amsmath,bm}
\usepackage{lineno}

\usepackage{cite,./mcite}
\usepackage{tikz}
\usepackage[symbol]{footmisc}

\newcommand{\dglap}{Gribov:1972ri,Lipatov:1974qm,Altarelli:1977zs,Dokshitzer:1977sg}

\def\qcdnum{{\sc QCDnum}}

\providecommand{\DOI}[1]{\href{http://dx.doi.org/#1}}

\def\updfevolv{{\sc uPDFevolv2}} 
\newcommand{\updfevolvversion}{2.6.0}

\newcommand{\as}{\ensuremath{\alpha_s}}

\begin{document}

\title{Parton Branching evolution in \updfevolv:\\
       NNLO spacelike evolution in the ZM-VFNS and\\
       NLO timelike evolution with heavy-flavour matching}
\author[1,2]{H.~Jung \thanks{hannes.jung@cern.ch}\orcidlink{0000-0002-2964-9845}}
\affil[1]{II. Institut f\"ur Theoretische Physik, Universit\"at Hamburg, Germany}
\affil[2]{Elementary Particle Physics, University of Antwerp, Belgium}
\date{November 2026}
\date{}
\begin{titlepage} 
\maketitle

\begin{abstract}
We describe two extensions of the Parton Branching (\PB) method as implemented in
\updfevolv: the next-to-next-to-leading order (NNLO) spacelike DGLAP evolution of parton densities
in the zero-mass variable-flavour-number scheme (ZM-VFNS), and the leading and next-to-leading order
(LO, NLO) timelike evolution of fragmentation functions, including the heavy-flavour matching
conditions at the quark thresholds.
We summarise the timelike \PB\ formalism --- momentum sum rule, Sudakov form factor, branching chain
and kinematics.
We give implementation details needed for the \PB\ formalism
to reproduce the semi-analytical solutions of \apfel\ for all
flavours, including heavy quarks at threshold.

This note supplements the \updfevolv\ manual~\cite{Hautmann:2014uua,Jung:2024uwc} and is 
a technical release note for \updfevolv\ version  \updfevolvversion .

\end{abstract}
\end{titlepage}

\tableofcontents

\section{Introduction}
The Parton Branching method~\cite{Hautmann:2017fcj,Hautmann:2017xtx} solves the DGLAP
equations~\cite{\dglap} by a Markov chain of
branchings, in which the non-resolvable and virtual contributions are resummed in the Sudakov form
factor and the resolvable emissions are restricted to $z<1-\epsilon$. At LO and NLO the collinear
part of the \PB\ densities agrees with the semi-analytical solution of \qcdnum~\cite{Botje:2010ay} and \apfel~\cite{Bertone:2017gds,Bertone:2013vaa}.
Since every branching is simulated, the method gives at the same time the transverse momentum
distribution of the partons.

This note describes two new developments in \updfevolv:
\begin{enumerate}
 \item \emph{Spacelike evolution at NNLO in the ZM-VFNS} (Sec.~\ref{sec:nnlo}).
   The extension to NNLO was outlined in Ref.~\cite{Zlebcik:2017suw}. It is not as
   straightforward as at LO and NLO, because in the VFNS the parton densities and $\alpha_s$
   are discontinuous at the heavy-quark thresholds~\cite{Buza:1996wv,Bierenbaum:2009zt}. Ref.~\cite{Zlebcik:2017suw}
   showed that these discontinuities can be included in the \PB\ evolution as $\delta$-function terms
   in the $\mu^2$-dependent splitting functions, which produce steps in the Sudakov form factor,
   and demonstrated agreement with \qcdnum\ for the gluon and the quarks.
      
 \item \emph{Timelike evolution at LO and NLO with heavy-flavour matching}
   (Secs.~\ref{sec:pb} and~\ref{sec:tl}). Fragmentation functions (FFs) obey DGLAP equations with
   timelike splitting functions. The \PB\ formulation of the timelike evolution is derived and 
    benchmarked against \apfel .
   In contrast to the parton densities, the FFs are discontinuous at the heavy-quark thresholds
   already at $\mathcal O(\alpha_s)$~\cite{Cacciari:2005ry}: the heavy-quark FF does not vanish at
   $\mu=m_h$ but is generated from the gluon FF. The implementation of this matching is described
   in Sec.~\ref{sec:tl}.
\end{enumerate}

\section{Spacelike and timelike evolution in the Parton Branching method \label{sec:pb}}
A parton density is denoted by $f_a(x,\mu^2)$ and the
fragmentation function of parton $a$ into a hadron $H$ by $D_a^H(x,\mu^2)$; the hadron index is
suppressed where it plays no role. The singlet combinations are
$f_\Sigma=\sum_{r=1}^{n_f}(f_{q_r}+f_{\bar q_r})$ and $D_\Sigma^H=\sum_{r=1}^{n_f}(D_{q_r}^H+D_{\bar q_r}^H)$,
and $[A\otimes B](x)\equiv\int_x^1\frac{\dd z}{z}A(z)B(x/z)$.

\subsection{Labelling convention and evolution equations}
For spacelike ($P^S_{ab}$) and timelike ($P^T_{ab}$) kernels alike, $P_{ab}(z)$ is the kernel of the
branching
\begin{equation}
 b\longrightarrow a\;(+\,c),
 \label{eq:branching-convention}
\end{equation}
with $z$ the momentum fraction of the daughter $a$ relative to the parent $b$. (Ref.~\cite{Furmanski:1980cm}
uses this labelling for spacelike kernels and the reversed one for timelike kernels.)
The two evolution equations differ in which index is summed: the PDF of $a$ is fed by all its
\emph{parents} $b$, the FF of the initiating parton $b$ by all its \emph{daughters} $a$,
\begin{align}
 \frac{\partial f_a(x,\mu^2)}{\partial\ln\mu^2}
 &= \frac{\alpha_s(\mu^2)}{2\pi}\sum_b\left[P^S_{ab}\otimes f_b\right](x,\mu^2)\qquad\text{(spacelike)},
 \label{eq:spacelike-dglap}\\
 \frac{\partial D_b^H(x,\mu^2)}{\partial\ln\mu^2}
 &= \frac{\alpha_s(\mu^2)}{2\pi}\sum_a\left[P^T_{ab}\otimes D_a^H\right](x,\mu^2)\qquad\text{(timelike)}.
 \label{eq:timelike-dglap}
\end{align}
At LO the Gribov--Lipatov relation~\cite{Gribov:1972ri,Lipatov:1974qm} gives
$P^{T,(0)}_{ab}=P^{S,(0)}_{ab}$ for the same branching. In the flavour-singlet sector, and
for flavour-symmetric quark and antiquark distributions, the corresponding equations are
\begin{align}
 \frac{\partial}{\partial\ln\mu^2}
 \begin{pmatrix} f_\Sigma\\ g\end{pmatrix}
 &=\frac{\alpha_s}{2\pi}
 \begin{pmatrix}
  P^S_{qq} & 2n_f P^S_{qg}\\
  P^S_{gq} & P^S_{gg}
 \end{pmatrix}\otimes
 \begin{pmatrix} f_\Sigma\\ g\end{pmatrix},
 \label{eq:singlet-spacelike}\\
 \frac{\partial}{\partial\ln\mu^2}
 \begin{pmatrix} D_\Sigma^H\\ D_g^H\end{pmatrix}
 &=\frac{\alpha_s}{2\pi}
 \begin{pmatrix}
  P^T_{qq} & 2n_f P^T_{gq}\\
  P^T_{qg} & P^T_{gg}
 \end{pmatrix}\otimes
 \begin{pmatrix} D_\Sigma^H\\ D_g^H\end{pmatrix}.
 \label{eq:singlet-timelike}
\end{align}
Thus, with the branching convention of Eq.~(\ref{eq:branching-convention}), the singlet
evolution matrices have the transposed off-diagonal structure; the factors $2n_f$ arise
from summing over quarks and antiquarks. Beyond LO the timelike kernels differ from the
spacelike ones and are taken from
Refs.~\cite{Curci:1980uw,Furmanski:1980cm,Floratos:1981hs,Stratmann:1996hn,Mitov:2006ic}.

\subsection{Momentum sum rule}
Momentum conservation in the incoming hadron for PDFs, and energy conservation for every initiating
parton for FFs, $\sum_H\int_0^1\dd z\,z\,D_a^H(z,\mu^2)=1$ (summed over all hadron species), lead to
the same condition on the kernels,
\begin{equation}
 \int_0^1\dd z\,z\left[P_{qq}(z)+P_{gq}(z)\right]=0,\qquad
 \int_0^1\dd z\,z\left[2n_f\,P_{qg}(z)+P_{gg}(z)\right]=0,
 \label{eq:kernel-sum-rule}
\end{equation}
valid order by order in $\alpha_s$ for the spacelike and for the timelike kernels; in the notation of
Eq.~(\ref{eq:branching-convention}), $\sum_a\int_0^1\dd z\,z\,P_{ab}(z)=0$ for any parent $b$, i.e.\ the
momentum fractions of all possible daughters add up to that of the parent. Beyond LO the sum over
daughters includes the $q\to\bar q$ and the pure-singlet $q_r\to q_s,\bar q_s$ channels.

\subsection{\PB\ evolution equations and Sudakov form factors\label{sec:PBevol}}
With the momentum-weighted distributions $\widetilde f_a=xf_a$ and $\widetilde D_a^H=xD_a^H$, the
virtual (and non-resolvable) contributions can be expressed through integrals of the real-emission
kernels with the help of the momentum sum rule, Eq.~(\ref{eq:kernel-sum-rule}), at any order in
$\alpha_s$ (for the spacelike case see~\cite{Hautmann:2017xtx}). The real-emission kernels $P^{R,S}_{ab}(z,\mu^2)$ and
$P^{R,T}_{ab}(z,\mu^2)$ are obtained from the full kernels by removing the $\delta(1-z)$ terms and
replacing $1/(1-z)_+$ by $1/(1-z)$; at NLO they contain a term of order $\alpha_s$. The real-emission
integrals are evaluated with the upper limit $1-\epsilon$; terms of $\mathcal O(\epsilon)$ are
neglected, and the collinear distributions are independent of $\epsilon$.\footnote{In
Refs.~\cite{Hautmann:2017xtx,Hautmann:2017fcj} this cut is formulated in terms of a soft-gluon resolution
scale $z_M=1-\epsilon$.} Relabelling the evolving parton as $a$, one obtains
\begin{align}
 \frac{\partial\widetilde f_a(x,\mu^2)}{\partial\ln\mu^2}
 &=\frac{\alpha_s(\mu^2)}{2\pi}\sum_b\int_x^{1-\epsilon}\dd z\,P^{R,S}_{ab}(z,\mu^2)\,
   \widetilde f_b\!\left(\frac{x}{z},\mu^2\right)
  -\frac{\alpha_s(\mu^2)}{2\pi}\,\mathcal V_a^S(\mu^2)\,\widetilde f_a(x,\mu^2),
 \label{eq:pb-spacelike}\\
 \frac{\partial\widetilde D_a^H(x,\mu^2)}{\partial\ln\mu^2}
 &=\frac{\alpha_s(\mu^2)}{2\pi}\sum_b\int_x^{1-\epsilon}\dd z\,P^{R,T}_{ba}(z,\mu^2)\,
   \widetilde D_b^H\!\left(\frac{x}{z},\mu^2\right)
  -\frac{\alpha_s(\mu^2)}{2\pi}\,\mathcal V_a^T(\mu^2)\,\widetilde D_a^H(x,\mu^2),
 \label{eq:pb-timelike}
\end{align}
with
\begin{equation}
 \mathcal V_a^{S,T}(\mu^2)\equiv\sum_b\int_0^{1-\epsilon}\dd z\,z\,P^{R,S/T}_{ba}(z,\mu^2),
 \label{eq:virtual-rate}
\end{equation}
where $b$ runs over all daughters of $a$, including the flavour and antiflavour multiplicities in the
singlet channel. The Sudakov form factors are
\begin{equation}
 \Delta_a^{S,T}(\mu^2,\mu_0^2)=\exp\left[-\int_{\mu_0^2}^{\mu^2}\frac{\dd\mu'^2}{\mu'^2}\,
 \frac{\alpha_s(\mu'^2)}{2\pi}\,\mathcal V_a^{S,T}(\mu'^2)\right],
 \label{eq:sudakov}
\end{equation}
and $\Delta_a(\mu^2,\mu_1^2)/\Delta_a(\mu^2,\mu_2^2)$ is the probability that parton $a$ does not branch
resolvably between $\mu_1^2$ and $\mu_2^2$. Integrating Eq.~(\ref{eq:pb-timelike}) gives
\begin{equation}
\begin{split}
 \widetilde D_a(x,\mu^2)={}&\Delta_a^T(\mu^2,\mu_0^2)\,\widetilde D_a(x,\mu_0^2)\\
 &+\sum_b\int_{\mu_0^2}^{\mu^2}\frac{\dd\mu'^2}{\mu'^2}\,\Delta_a^T(\mu^2,\mu'^2)\,
 \frac{\alpha_s(\mu'^2)}{2\pi}\int_x^{1-\epsilon}\dd z\,P^{R,T}_{ba}(z,\mu'^2)\,
 \widetilde D_b\!\left(\frac{x}{z},\mu'^2\right),
\end{split}
 \label{eq:pb-integral-timelike}
\end{equation}
and analogously for the spacelike case. An explicit LO example of how the momentum sum rule turns
the virtual term into the Sudakov form factor is given in Appendix~\ref{sec:timelike-pb-form}.
Comparing Eqs.~(\ref{eq:pb-spacelike}) and~(\ref{eq:pb-timelike}) we observe:
\begin{itemize}
 \item \emph{Sudakov form factor:} in both cases it is built from the daughters of the evolving
   parton, $\sum_b\int z P_{ba}$, as a consequence of the same momentum sum rule. At LO the spacelike
   and timelike Sudakov form factors are identical.
 \item \emph{Real emission:} the PDF of $a$ is fed by its parents, $\sum_bP^{S}_{ab}$; the FF of $a$ is
   fed by its daughters, $\sum_bP^{T}_{ba}$, i.e.\ by the same kernel combination as in the Sudakov
   form factor.
\end{itemize}

\subsection{Branching chain}\label{sec:chain}
Iterating Eq.~(\ref{eq:pb-integral-timelike}), or its spacelike analogue, generates a chain of
branchings starting at $\mu_0$, in which a parton $b$ evolves with $\Delta_b$ to the scale of the next
branching, where it is replaced by a parton $a$ with $x\to zx$. In the spacelike case $a$ is the
physical daughter of $b$, and the chain follows the physical branching direction. In the timelike
case $a$ is the physical \emph{parent} of $b$: the chain runs from the hadron towards the initiating
parton.

In the timelike chain the scale of the next branching is generated with $\Delta^T_b$, which is
normalised to the daughters of $b$, while the new parton $a$ is selected among the parents of $b$.
Each branching is therefore multiplied by a factor
\begin{equation}
 w_b(\mu'^2)=\frac{\sum_a\int_0^{1-\epsilon}\dd z\,z\,P^{R,T}_{ba}(z,\mu'^2)}
                  {\sum_c\int_0^{1-\epsilon}\dd z\,z\,P^{R,T}_{cb}(z,\mu'^2)} .
 \label{eq:timelike-weight}
\end{equation}
This factor differs from one only for the $q\leftrightarrow g$ transitions, and is identically one in the spacelike case.

\subsection{Kinematics and ordering}\label{sec:kin}
For the collinear distributions the relation between the evolution scale and the kinematics of the
branching plays no role, as long as $\alpha_s$ is evaluated at the evolution scale and a fixed cut
$1-\epsilon$ is used. For transverse momentum dependent distributions, and whenever $\alpha_s$ depends
on the transverse momentum of the emission, it fixes the transverse momentum \qt\ of every emitted
parton. With angular ordering~\cite{Ciafaloni:1987ur,Catani:1989yc,Catani:1989sg,Marchesini:1994wr}
the evolution scale is $\mu'=E_b\theta$, with $\theta$ the ordered angle, and
\begin{equation}
 \qt=(1-z)\,\mu'\quad\text{(spacelike)},\qquad\qquad
 \qt=z(1-z)\,\mu'\quad\text{(timelike)}.
 \label{eq:qt}
\end{equation}
In the spacelike case the ordered angle is that of the emitted parton $c$ relative to the incoming
hadron; in the timelike case colour coherence orders the opening angle of the pair
$a+c$~\cite{Marchesini:1983bm,Marchesini:1987cf}. Both relations coincide for soft emissions,
$z\to1$. Angular ordering is the default in \updfevolv; in addition, \qt\ ordering ($\qt=\mu'$) is
available in the new version.

\section{Spacelike evolution at NNLO: variable-flavour thresholds}\label{sec:nnlo}
\subsection{Matching conditions}
Below $\mu^2=m_h^2$ ($h=c,b,t$) there are $n_f$ active flavours, above it $n_f+1$. The physical
cross section is continuous across the threshold only if the parton densities are matched
with the operator matrix elements (OMEs) of Ref.~\cite{Buza:1996wv,Bierenbaum:2009zt}. In this section we use the
normalisation of Ref.~\cite{Buza:1996wv,Bierenbaum:2009zt}, $\bar{a}_s=\alpha_s^{(n_f+1)}(m_h^2)/4\pi$, and keep its names of
the OMEs, in which the index $H$ denotes the heavy quark. With the on-shell mass $m_h$, and all
densities on the right-hand side being the $n_f$-flavour ones at $\mu^2=m_h^2$, the
$\mathcal O(\bar{a}_s)$ terms vanish, and through $\mathcal O(\bar{a}_s^2)$
\begin{align}
 g^{(n_f+1)} &= g + \bar{a}_s^2\left[A_{gq,H}\otimes\Sigma + A_{gg,H}\otimes g\right],\label{eq:mg}\\
 (q_k+\bar q_k)^{(n_f+1)} &= (q_k+\bar q_k)+\bar{a}_s^2\,A^{\rm NS}_{qq,H}\otimes(q_k+\bar q_k),
   \qquad k=1,\dots,n_f,\label{eq:mq}\\
 (h+\bar h) &= \bar{a}_s^2\left[\tilde A^{\rm PS}_{Hq}\otimes\Sigma+\tilde A_{Hg}\otimes g\right],
 \label{eq:mh}
\end{align}
where $\Sigma=\sum_{k=1}^{n_f}(q_k+\bar q_k)$, and $\tilde A^{\rm PS}_{qq,H}$ and
$\tilde A_{qg,H}$ vanish at this order~\cite{Buza:1996wv,Bierenbaum:2009zt}.  

The kernels are given in Appendix~B of Ref.~\cite{Buza:1996wv,Bierenbaum:2009zt} at $\mu=m_h$: $\tilde A_{Hg}$
(B.3), $\tilde A^{\rm PS}_{Hq}$ (B.1), $A_{gq,H}$ (B.5), $A_{gg,H}$ (B.7) and $A^{\rm NS}_{qq,H}$ (B.4).
For $\tilde A_{Hg}$ we use the compact parametrisation of Vogt~\cite{Vogt:2004ns}.

\subsection{Implementation in the branching Monte Carlo}\label{sec:impl}
Following Ref.~\cite{Zlebcik:2017suw}, the kernels are included as part of the
$\mu^2$-dependent real-emission kernels,
\begin{equation}
 \frac{\alpha_s(\mu^2)}{2\pi}\,P^{R,S}_{ab}(z,\mu^2)\;\to\;
 \frac{\alpha_s(\mu^2)}{2\pi}\,P^{R,S}_{ab}(z,\mu^2)
 +\bar{a}_s^2(m_h^2)\,A_{ab}(z)\,\delta(\ln\mu^2-\ln m_h^2),
 \label{eq:thr}
\end{equation}
so that the $\delta$-function produces a step of the Sudakov form factor, Eq.~(\ref{eq:sudakov}), at
each threshold. Because the kernels satisfy the sum rules, the threshold terms have the
same real/virtual structure as the regular kernels, and the step of the Sudakov form factor is the
corresponding virtual contribution. The virtual terms of $A_{ab}$ at $z=1$ are not used, in the same
way as for $P^{R,S}$ (since included via the Sudakov form factor). When the Markov chain reaches $\mu^2=m_h^2$, the branching type and the value of
$z$ are selected according to the kernels $A_{ba}$ instead of $P^{R,S}_{ba}$.

The assignments are (parent $\to$ daughter, kernel), with the labelling of
Eq.~(\ref{eq:branching-convention}):
\begin{center}
\begin{tabular}{lll}
$g\to h,\bar h$ & $\tilde A_{Hg}$ & sign of daughter random, probability $\tfrac12$ each\\
$q_i,\bar q_i\to h,\bar h$ & $\tilde A^{\rm PS}_{Hq}$ & weight $\tfrac12$ to $h$ and $\tfrac12$ to $\bar h$ per parent\\
$g\to g$ & $A_{gg,H}$ & regular + soft part\\
$q_i\to g$, $\bar q_i\to g$ & $A_{gq,H}$ &\\
$q_i\to q_i$, $\bar q_i\to\bar q_i$ & $A^{\rm NS}_{qq,H}$ & regular + soft part, full weight\\
$q_i\to\bar q_i$ & $0$ & $A^{\rm NS,+}_{qq,H}=A^{\rm NS,-}_{qq,H}$ at $\mathcal O(\bar{a}_s^2)$\\
\end{tabular}
\end{center}
The coupling in the threshold kernels is $\bar{a}_s(m_h^2)$ in the $(n_f+1)$-flavour scheme.

\section{Timelike evolution at NLO: heavy-flavour thresholds}\label{sec:tl}
\subsection{Matching conditions}
In the timelike variable-flavour-number scheme the heavy-quark FFs are generated dynamically: below
the threshold $\mu=m_h$ ($h=c,b$) they vanish, and at $\mu=m_h$ they are fixed by the
$\overline{\rm MS}$ matching conditions of Ref.~\cite{Cacciari:2005ry}. For thresholds placed at the
quark masses they read, through $\mathcal O(\alpha_s)$,
\begin{align}
 D_{q_k}^{(n_f+1)}(x,m_h^2) &= D_{q_k}(x,m_h^2), \qquad k=1,\dots,n_f, \label{eq:tlq}\\
 D_g^{(n_f+1)}(x,m_h^2) &= D_g(x,m_h^2), \label{eq:tlg}\\
 D_h(x,m_h^2) = D_{\bar h}(x,m_h^2)
   &= \frac{\alpha_s(m_h^2)}{2\pi}\int_x^1\frac{\dd z}{z}\,K(z)\,D_g\!\left(\frac{x}{z},m_h^2\right),
   \label{eq:hqmatching}\\
 K(z) &= C_F\,\frac{1+(1-z)^2}{z}\,\bigl[-1-2\ln z\bigr],
 \label{eq:K}
\end{align}
with the $n_f$-flavour FFs on the right-hand side; $\alpha_s$ is continuous across the threshold at
this order. The kernel describes the emission of a gluon off the heavy quark, $h\to h\,g$, i.e.\ it
belongs to the transition $P_{gh}$ in the labelling of Eq.~(\ref{eq:branching-convention}).
For a general matching scale the kernel of Ref.~\cite{Cacciari:2005ry} is
$C_F\frac{1+(1-z)^2}{z}\bigl[\ln(\mu^2/m_h^2)-1-2\ln z\bigr]$; for thresholds placed at the quark
masses the logarithm vanishes, and Eq.~(\ref{eq:K}) remains.

The structure of the timelike matching differs from the spacelike one in several respects:
\begin{itemize}
 \item The heavy-quark FF is non-zero at threshold already at $\mathcal O(\alpha_s)$, whereas the
   spacelike matching starts at $\mathcal O(\alpha_s^2)$.
 \item The light-quark and gluon FFs are continuous at $\mathcal O(\alpha_s)$.
 \item At this order, the heavy-quark matching condition involves the gluon FF only,
   as shown in Eq.~(\ref{eq:hqmatching}). In the branching convention of
   Eq.~(\ref{eq:branching-convention}), the corresponding kernel is $P_{gh}$:
   the initiating heavy quark has a gluon daughter in the timelike branching description.
\end{itemize}

\begin{figure}[htb]
 \centering
 \includegraphics[width=0.45\textwidth]{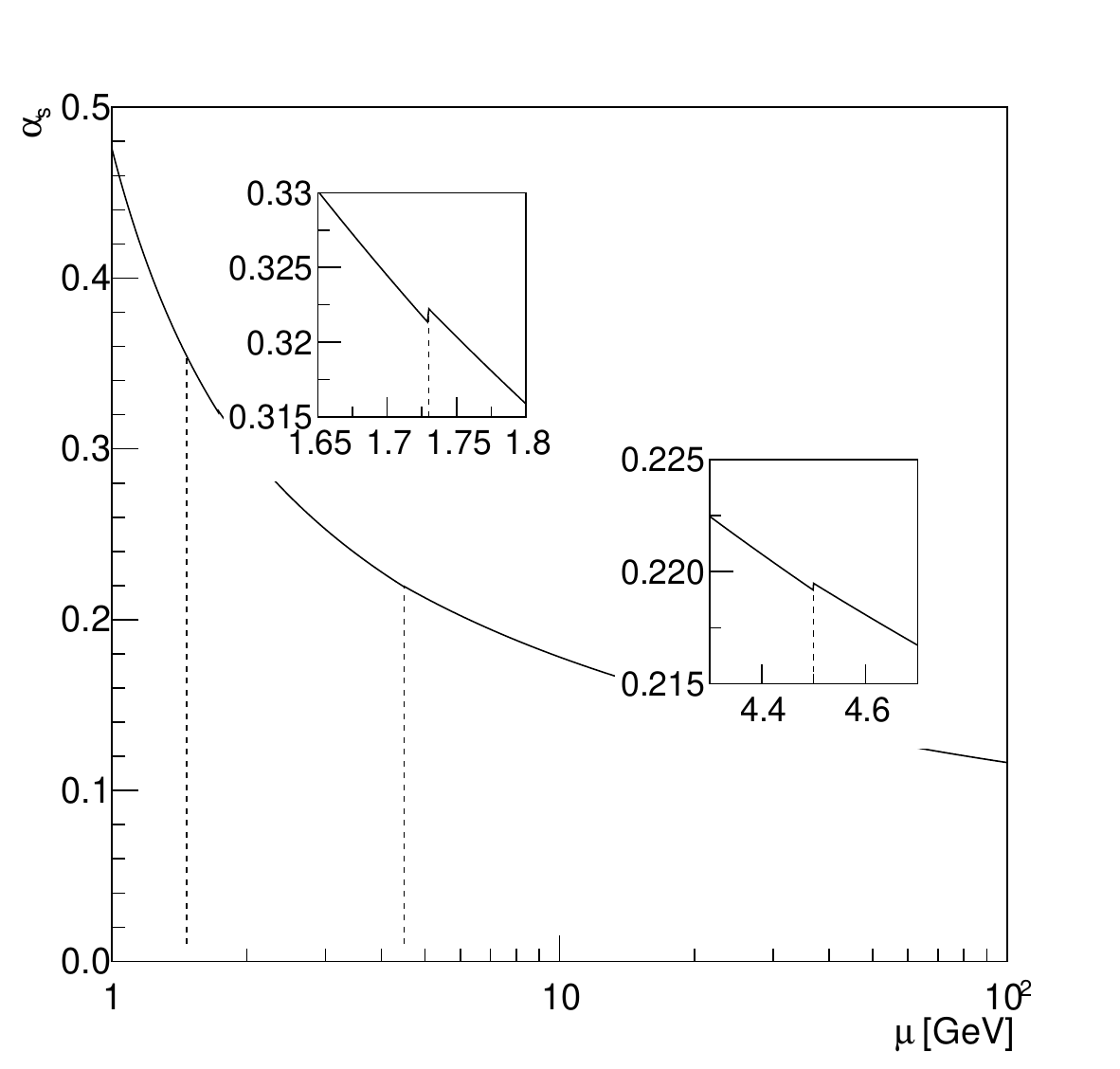}
 \includegraphics[width=0.45\textwidth]{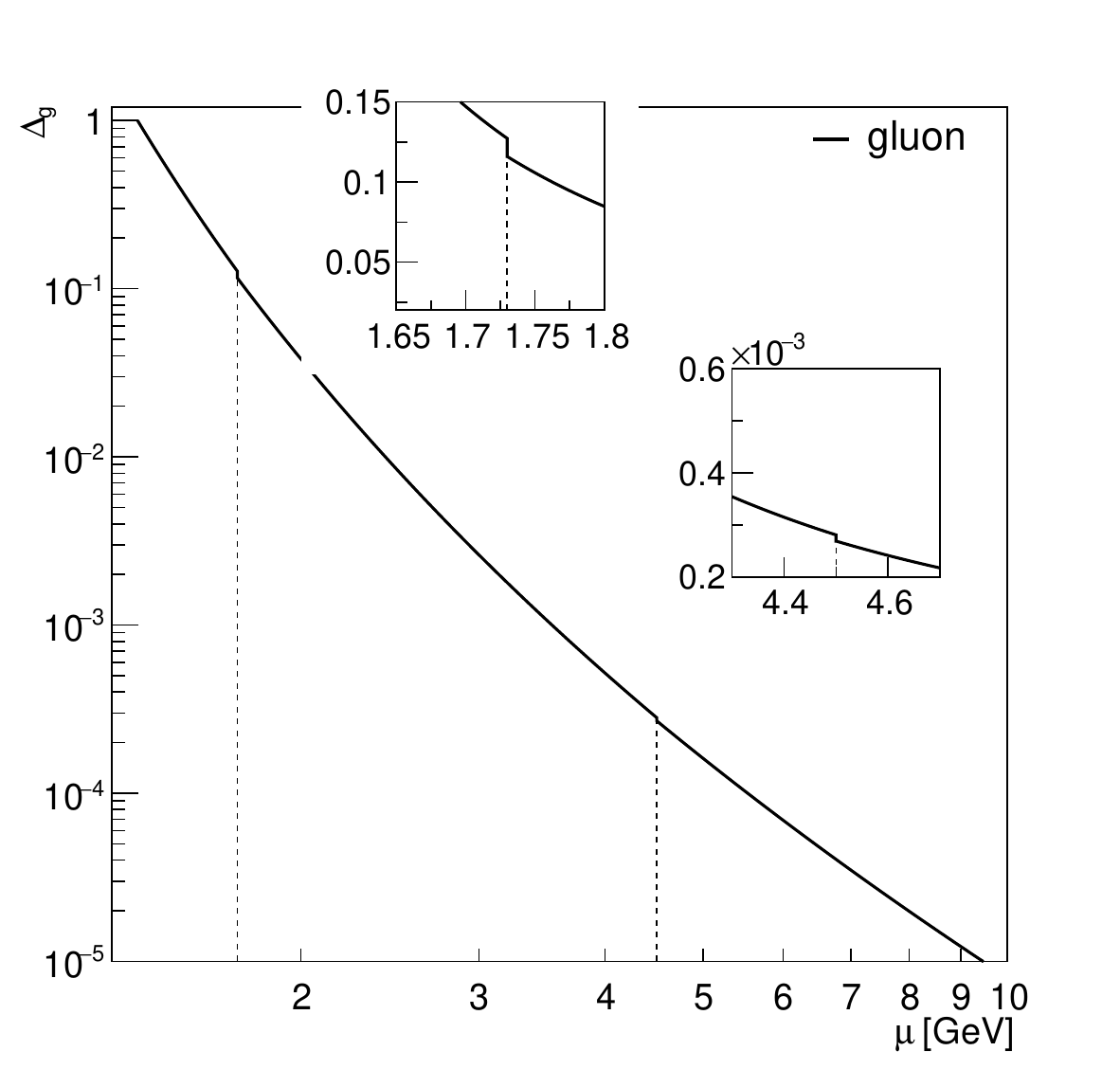}
 \caption{$\as$ at three-loop (left) and  the NNLO Sudakov form factor (right) of the gluon
 as a function of $\mu$. The dotted lines mark $m_c$ and $m_b$; the insets show the steps generated by \as\ at three-loops and
 the $\delta(\ln\mu^2-\ln m_h^2)$ terms of Eq.~(\ref{eq:thr}). 
 }
 \label{fig:sud}
\end{figure}

\subsection{Implementation in the branching Monte Carlo}\label{sec:impltl}
In contrast to the spacelike case, Eq.~(\ref{eq:hqmatching}) cannot be included through the Sudakov
form factor: it is a real contribution without a corresponding virtual term, and a discontinuity of
the gluon Sudakov form factor at $m_h$ would deplete $D_g$, in contradiction to its continuity,
Eq.~(\ref{eq:tlg}). The Sudakov form factors are therefore left unchanged, and the matching is applied
as a separate step in the branching chain of Sec.~\ref{sec:chain}:
\begin{tolerant}{5000}
\begin{enumerate}
 \item When the next evolution scale of a gluon would cross a threshold, the chain is stopped at
   exactly $\mu=m_h$. Since the Sudakov form factor factorises,
   $\Delta(\mu^2,\mu_0^2)=\Delta(\mu^2,m_h^2)\,\Delta(m_h^2,\mu_0^2)$, the evolution can be restarted
   from $m_h$ without bias.
 \item At $m_h$ the gluon either continues as a gluon, with unit rate, since $D_g$ is continuous,
   or turns into a heavy quark or antiquark according to Eq.~(\ref{eq:hqmatching}). Since the matching
   does not deplete the gluon FF, the continuation and matching contributions cannot be interpreted
   as mutually exclusive physical probabilities. One of the two possibilities is therefore selected
   with an auxiliary algorithmic probability $p$, and the event weight compensates for this sampling choice: with probability $1-p$ the chain continues as a gluon with weight
   $1/(1-p)$; with probability $p$ it is turned into a heavy quark or antiquark (with equal probability),
   carrying the momentum fraction $z\,x_g$. The value of $z$ is sampled uniformly in $\ln z$, and the
   weight is $\propto\alpha_s(m_h^2)\,z\,K(z)/p$, where the factor $z$ is the Jacobian
   $\dd z=z\,\dd\ln z$ of the sampling. On average this reproduces $D_g$ and Eq.~(\ref{eq:hqmatching}) exactly,
   independently of $p$; in \updfevolv\ $p=0.2$ is used.
 \item The heavy quark evolves from $m_h$ with the $(n_f+1)$-flavour splitting functions. Its
   transverse momentum at the threshold step is obtained with the same kinematic prescription,
   Eq.~(\ref{eq:qt}), as for an ordinary branching at the scale $m_h$.
\end{enumerate}
\end{tolerant}

\section{Benchmark comparison:  \apfel}
\subsection{Spacelike NNLO evolution}
The \PB\ evolution is compared to the semi-analytical NNLO VFNS solution of \apfel\
with the same starting distributions, the same NNLO splitting functions~\cite{Moch:2004pa,Vogt:2004mw},
the same thresholds ($m_c=1.73$\,GeV, $m_b=4.5$\,GeV, $m_t=175$\,GeV), and the same $\alpha_s$
evolution with the corresponding discontinuities. The cut on the real emissions is
$\epsilon=10^{-6}$.
\begin{figure}[htb]
 \centering
 \includegraphics[width=0.4\textwidth]{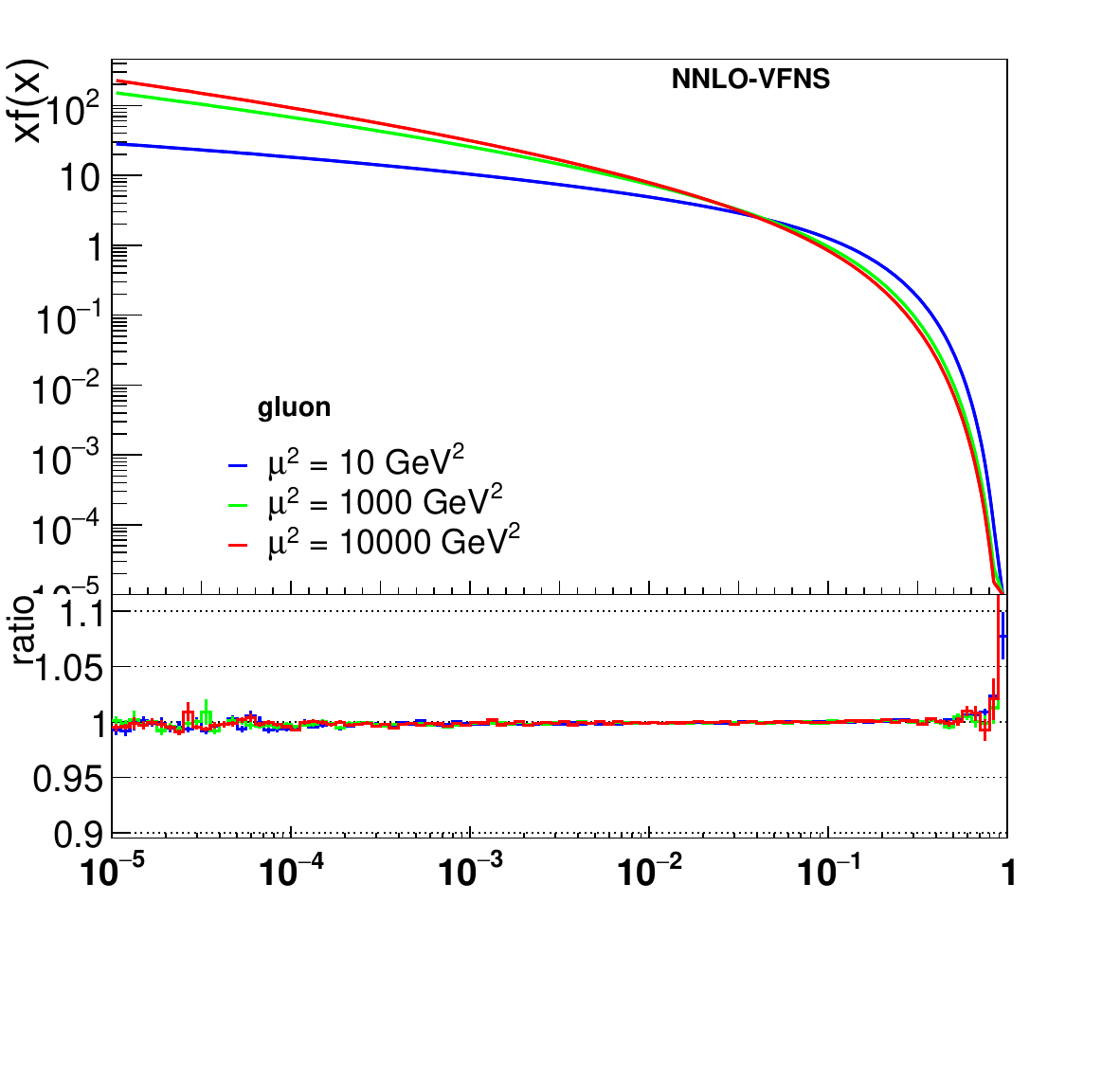}
 \includegraphics[width=0.4\textwidth]{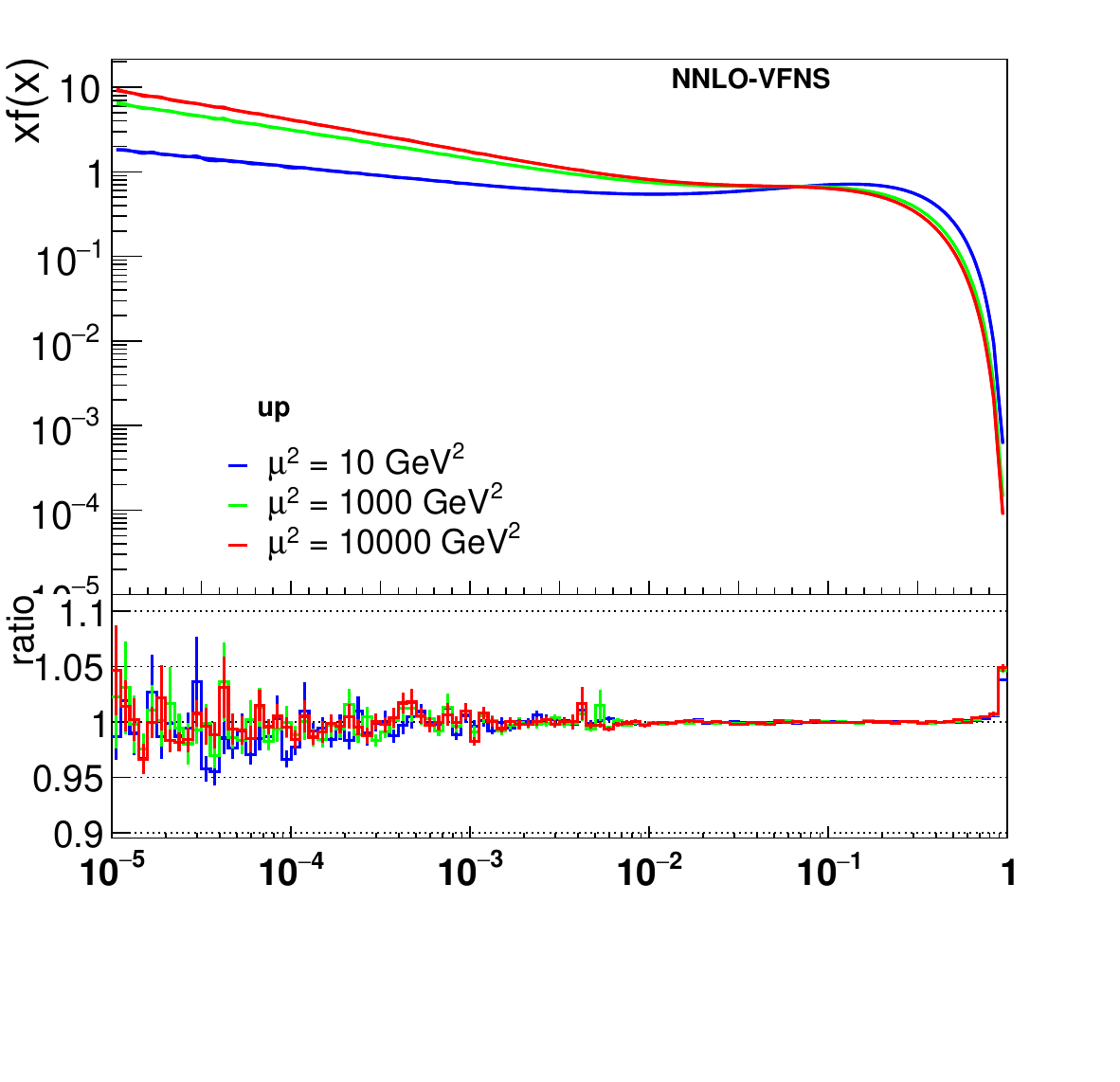}
 \vspace*{-1cm}
  \caption{$xg(x,\mu^2)$ and $xu(x,\mu^2)$ from the \PB\ method (histograms) and \apfel
 (lines) at NNLO for $\mu^2=10,10^3,10^5$\,GeV$^2$. The lower panels show the ratio to \apfel .
 See Fig.~3 of Ref.~\protect\cite{Zlebcik:2017suw}. }
 \label{fig:gu}
\end{figure}
\begin{figure}[htb]
 \centering
 \includegraphics[width=0.4\textwidth]{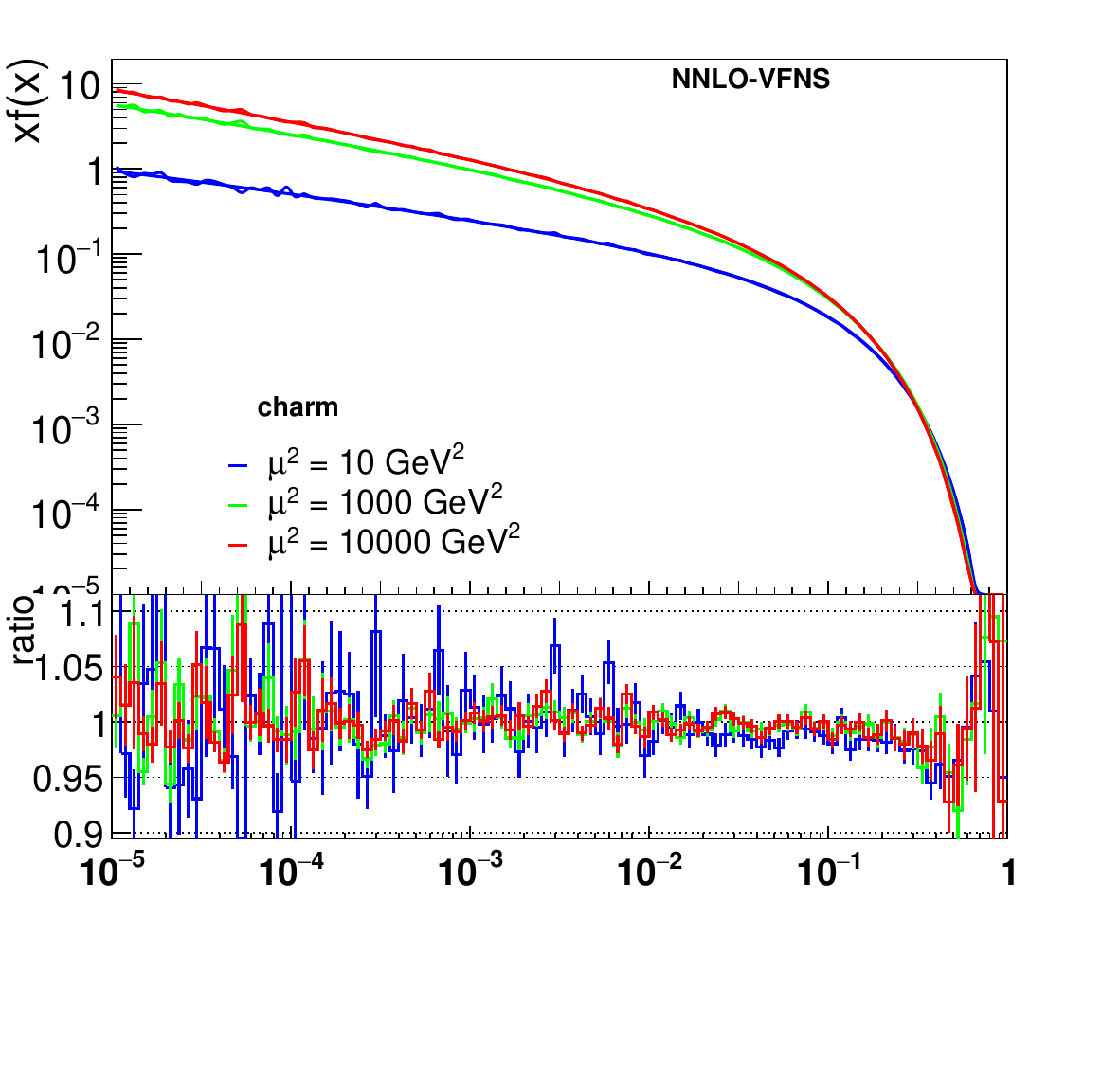} \includegraphics[width=0.4\textwidth]{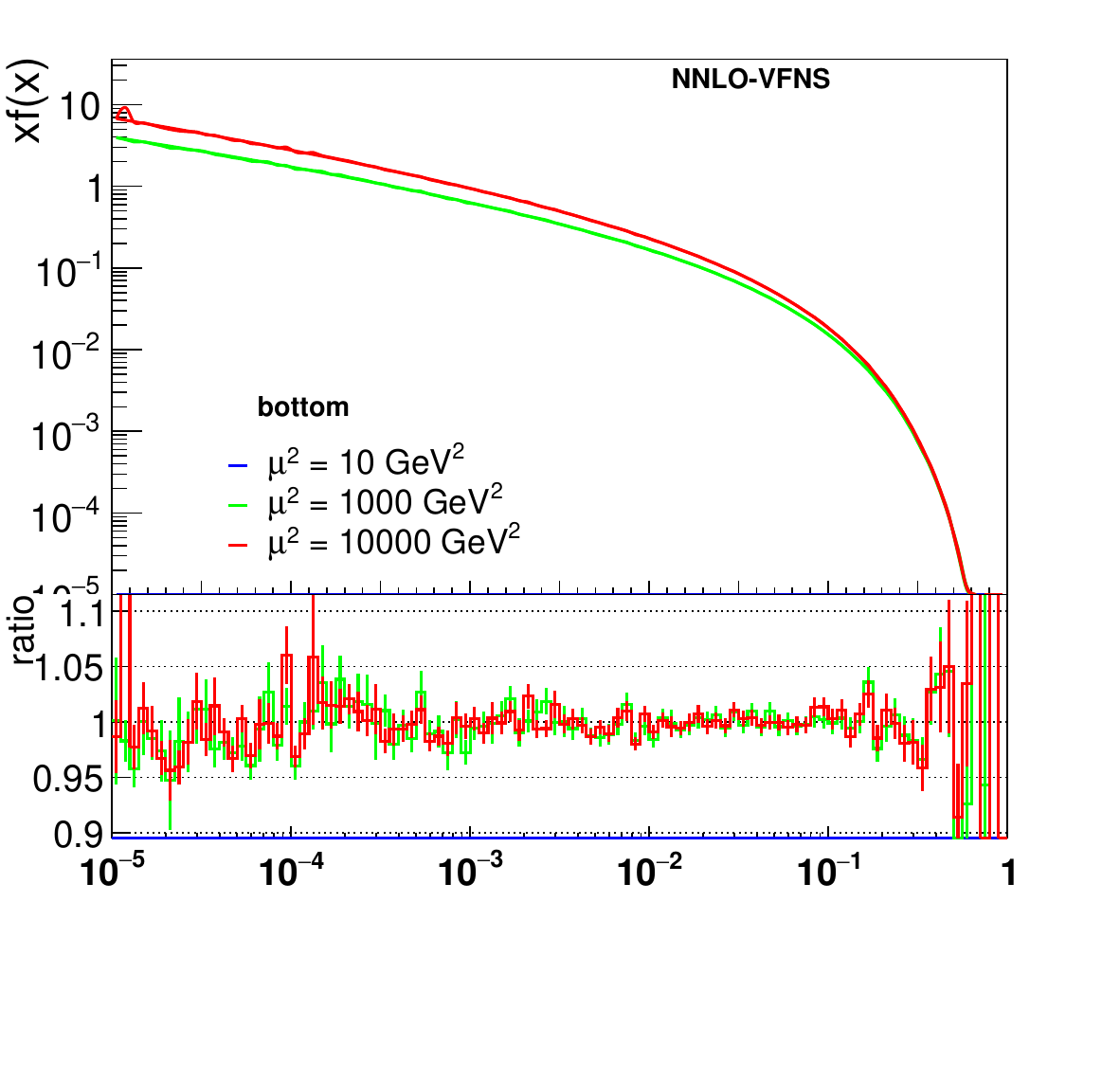}
 \vspace*{-1cm} \caption{As Fig.~\ref{fig:gu} for 
 $xc$ and $xb$. The heavy quarks probe the pure-singlet and gluon kernels at threshold.}
 \label{fig:qbar}
\end{figure}

Figure~\ref{fig:sud} shows \as\ at three-loops and the Sudakov form factor with the discontinuities at the
heavy-quark thresholds.
Figures~\ref{fig:gu} and \ref{fig:qbar} show the comparison with \apfel\ for the gluon, the
up quark, and the charm and bottom densities.

\subsection{Timelike NLO evolution}
The timelike \PB\ evolution is compared with the \apfel\ reference calculation, using the
heavy-flavour matching conditions of Ref.~\cite{Cacciari:2005ry} and the timelike evolution
reference results of Ref.~\cite{Bertone:2015cwa}. The starting scale is $Q_0=1.37$\,GeV, with
pion starting distributions taken from Ref.~\cite{Albino:2005me},
\begin{equation}
\begin{aligned}
 xD_g(x,Q_0^2) &= 429\,x^2(1-x)^{5.82}, &\qquad xD_u(x,Q_0^2) &= 0.447\,x^{-1.58}(1-x)^{1.01},\\
 xD_d(x,Q_0^2) &= 0.833\,x^{-1.17}(1-x)^{1.39}, &\qquad xD_s(x,Q_0^2) &= 0.519\,x^{-0.365}(1-x)^{1.96},
\end{aligned}
\end{equation}
with all other distributions vanishing at $Q_0$, $\asmz=0.118$ evolved at NLO, and NLO splitting
functions. 

\begin{figure}[htb]
 \centering
 \includegraphics[width=0.4\textwidth]{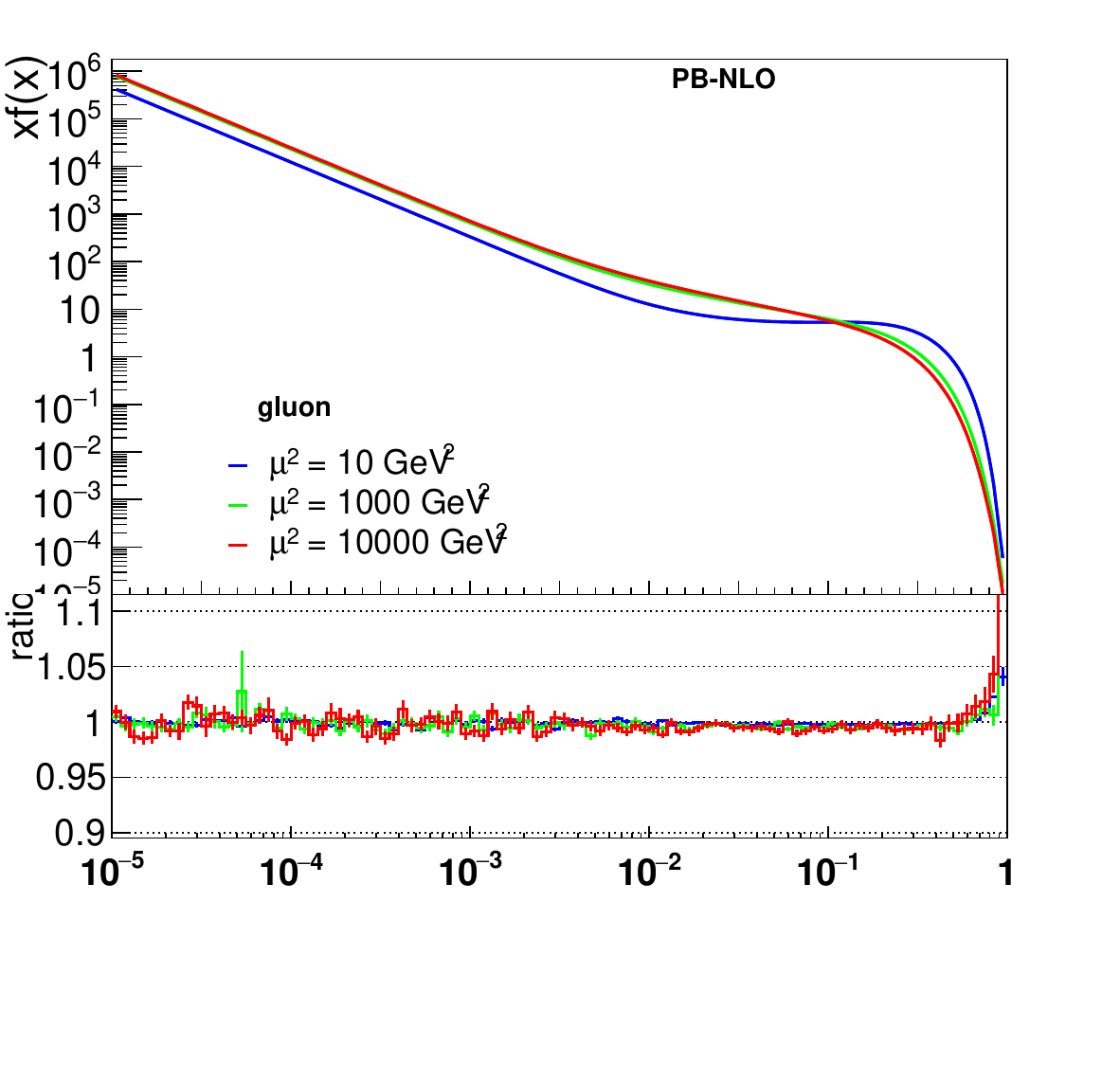} \includegraphics[width=0.4\textwidth]{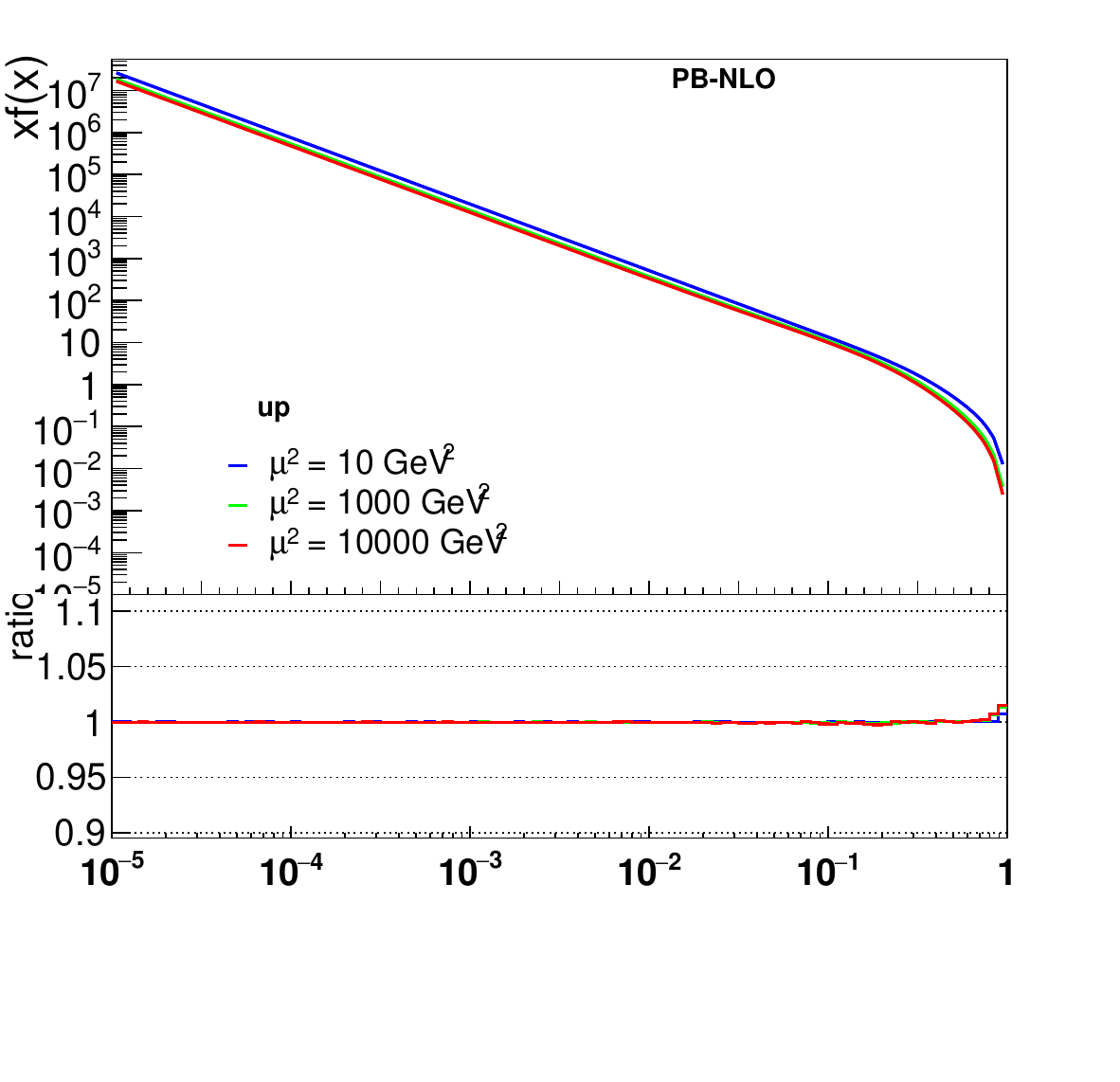}
 \includegraphics[width=0.4\textwidth]{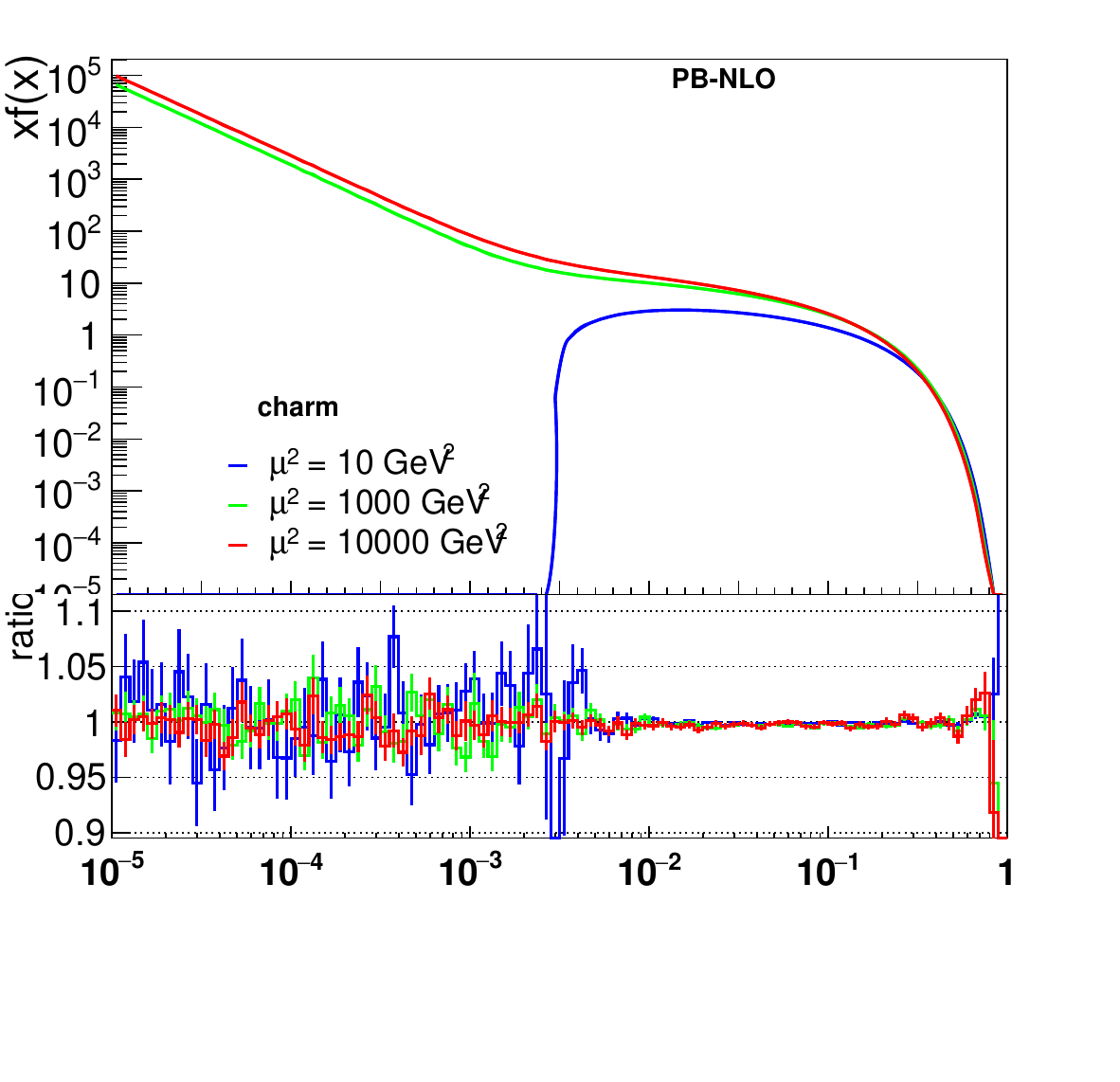} \includegraphics[width=0.4\textwidth]{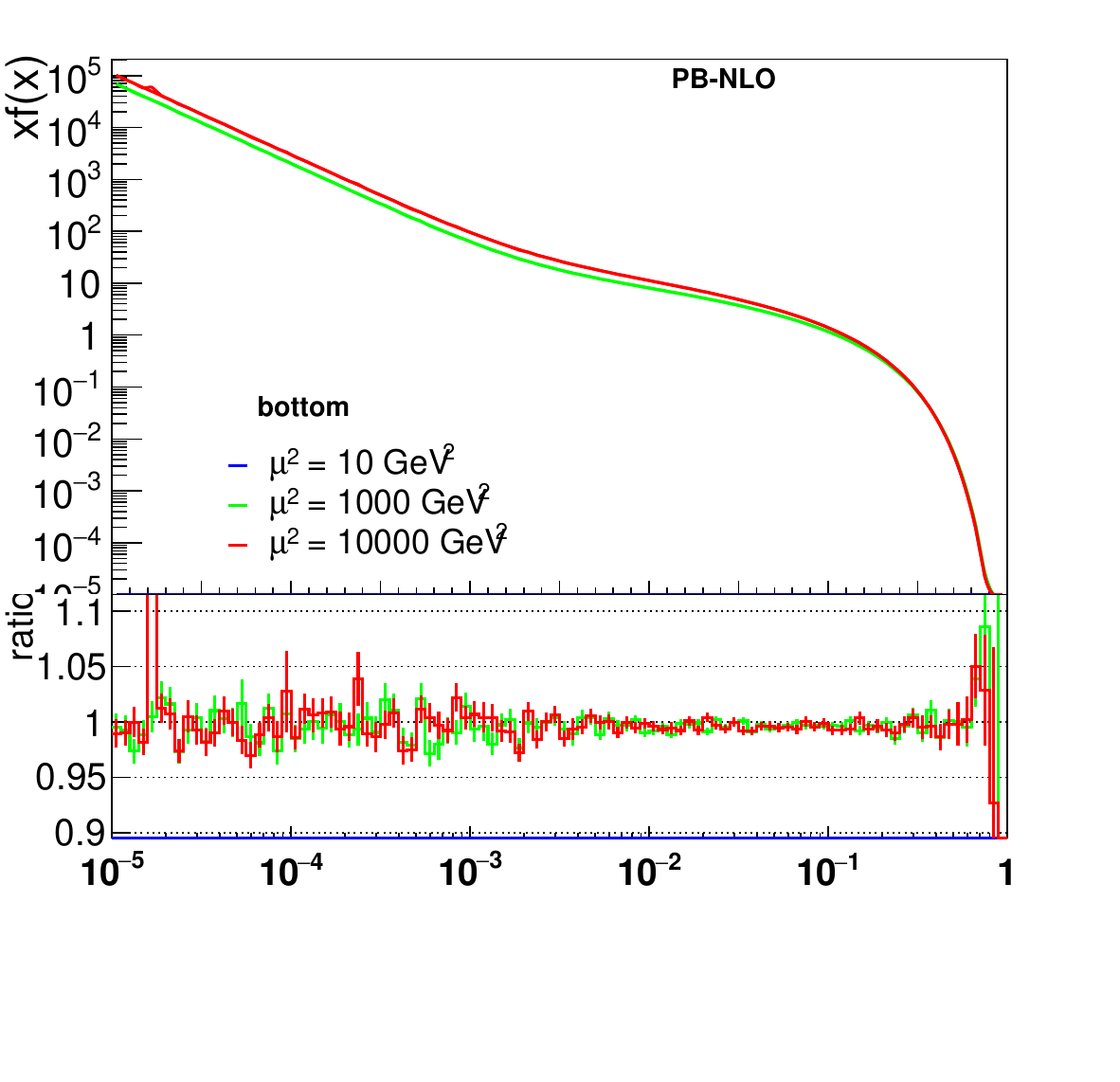}
  \vspace*{-1cm}
 \caption{Fragmentation functions of gluon and up quark  and
 of charm and bottom quark (lower row) from timelike evolution at NLO with the \PB\ method and with
 \apfel, using the same initial conditions. The heavy-quark FFs probe the threshold kernel $K$ of
 Eq.~(\protect\ref{eq:K}).}
 \label{fig:tl}
\end{figure}

Figure~\ref{fig:tl} shows the comparison. The \PB\ results agree well with \apfel\ for all
flavours, including the heavy-quark distributions above threshold. The remaining fluctuations,
most visible for charm at small $x$, are due to the finite Monte Carlo statistics.

\section{Usage in \updfevolv}
The source code of \updfevolv\ and this manual can be found under:\\
\verb+https://updfevolv.hepforge.org/+

\begin{tolerant}{9000}
All parameters can be set via the steering cards:
\begin{defl}{123456789000000000}
\item[]
\item[{\tt  Iorder\_alphas=2}]   order in perturbation theory for splitting functions as well as \as  \\ 
 LO: {\tt Iorder\_alphas = 1}\\ NLO: {\tt Iorder\_alphas=2}\\  NNLO: {\tt Iorder\_alphas=3}.  
\item[{\tt  I\_Timelike=0}]  switch for spacelike/timelike evolution
\item {\tt I\_Timelike=0}: spacelike evolution
\item {\tt I\_Timelike=1}: timelike evolution
\item[{\tt  Iqordering=2}]   ordering definition to calculate scale in $\alpha_s$, the transverse momentum \qt\ and \zM. Please note that the calculation of \qt\  changes for timelike evolution, $\qt = z(1-z)q'$ for angular ordering (Eq.~(\ref{eq:qt}))
\item {\tt Iqordering=0}: Angular ordering with  $\alpha_s(\qt)$, $\qt = (1-z) q^{\prime}$ and  $\zM=z_{\rm{dyn}} = 1-q_0/q'$
\item {\tt Iqordering=1}: Virtuality ordering with $\alpha_s(q)$, $\qt = \sqrt{(1-z)} q^{\prime}$  and  $\zM=z_{\rm{dyn}} = 1-(q_0/q')^2$
\item {\tt Iqordering=2}:  Angular ordering with  $\alpha_s(q)$, $\qt = (1-z) q^{\prime}$ and   fixed \zM
\item {\tt Iqordering=3}:  Virtuality ordering with $\alpha_s(q)$, $\qt = \sqrt{(1-z)} q^{\prime}$ and  $\zM=z_{\rm{dyn}} = 1-(q_0/q')^2$ 
\item {\tt Iqordering=4}: Angular ordering with  $\alpha_s(\qt)$, $\qt = (1-z) q^{\prime}$ and   fixed \zM
\item {\tt Iqordering=5}: \qt\ ordering ( $\qt = q$)  with $\alpha_s(q)$ and fixed \zM  
\end{defl}
\end{tolerant}

\section{Summary}
The \PB\ implementation in \updfevolv\ has been extended to NNLO spacelike evolution in the
ZM-VFNS and to LO and NLO timelike evolution with heavy-flavour matching.

For spacelike evolution, the NNLO threshold matching coefficients are implemented
as $\delta(\ln\mu^2-\ln m_h^2)$ contributions to the scale-dependent kernels. They generate the
corresponding steps in the Sudakov form factor. The resulting gluon, light-quark and heavy-quark
densities agree with the semi-analytical \apfel\ results.

For timelike evolution, the Sudakov form factor is constructed from the daughters of the evolving
parton, while the branching chain runs from the hadron towards the initiating parton and uses the
transposed kernel structure. The heavy-flavour matching  starts at
$\mathcal O(\alpha_s)$ and contributes to the heavy-quark FF through the gluon FF. Since this
matching has no corresponding virtual term, it is implemented as a separate threshold step in the
branching chain, with an auxiliary sampling probability and a compensating event weight. The
resulting fragmentation functions agree well with the \apfel\ reference calculation.

\section*{Acknowledgements}
We thank V.~Bertone for discussions on timelike evolution.

Generative AI tools were used during the development of the implementation and the preparation
of this note. These included Claude (Anthropic; Claude Sonnet 5 and Claude Opus 5.5),
ChatGPT (OpenAI), and Qwen. They were used for initial studies, assistance with implementation
and documentation, and cross-checks. The scientific results, equations, references and conclusions
were critically reviewed by the author, who takes full responsibility for the content of this note.

\appendix
\section{Explicit LO example}
\label{sec:timelike-pb-form}
\begin{tolerant}{5000}
As an illustration of the general derivation of Sec.~\ref{sec:PBevol}, we show
explicitly for quark and gluon how the momentum sum rule turns the virtual term into the
Sudakov term. For a kernel of pure plus-prescription type, $P=[P^R]_+$, as $P_{qq}$ at LO, the
$z$-moment is
\begin{equation}
\int_0^1 dz \, z \left[P^R(z)\right]_+ = \int_0^1 dz \left[ \,z {P}^{R}(z) - {P}^{R}(z) \right] .
\label{eq:plus-z-moment}
\end{equation}
Applying the momentum sum rule $\int_0^1 dz\,\left[z P_{qq}(z) + z P_{gq}(z)\right] = 0$, Eq.~\eqref{eq:kernel-sum-rule}, the evolution equation of the quark FF can be written in terms of the real-emission timelike kernels $P^{R,T}_{ba}$ as
\begin{eqnarray}
 \frac{\partial\, xD_q(x,\mu^2)}{\partial\ln\mu^2}
& = &
 \frac{\as}{2\pi}
  \int_x^{1} \frac{dz}{z} \left[ z {P}^{R,T}_{qq} \left(z\right) \frac{x}{z}{D_q}\left({\frac{x}{z}}, \mu^{2}\right) +   {z P}^{R,T}_{gq} \left(z\right) \frac{x}{z}{D_g}\left({\frac{x}{z}}, \mu^{2}\right) \right]  \nonumber \\
  &   &  - \frac{\as}{2\pi}
    xD_q(x,\mu^2)\int_0^{1} dz { P}^{R,T}_{qq}
   \left(z\right) \nonumber \\
  & & -  \frac{\as}{2\pi}  xD_q(x,\mu^2) \left[ \int_0^1 dz \, z { P}_{qq}(z) + z P_{gq}(z) \right]  \nonumber \\
& = &
 \frac{\as}{2\pi}
  \int_x^{1} \frac{dz}{z} \left[ z {P}^{R,T}_{qq} \left(z\right) \frac{x}{z}{D_q}\left({\frac{x}{z}}, \mu^{2}\right) +   {z P}^{R,T}_{gq} \left(z\right) \frac{x}{z}{D_g}\left({\frac{x}{z}}, \mu^{2}\right) \right]  \nonumber \\
  &   &  - \cancel{\frac{\as}{2\pi}
    xD_q(x,\mu^2)\int_0^{1} dz {P}^{R,T}_{qq}
   \left(z\right) } \nonumber \\
  & & -  \frac{\as}{2\pi}  xD_q(x,\mu^2) \left[ \int_0^1 dz \, z { P}^{R,T}_{qq}(z) - \cancel{{ P}^{R,T}_{qq}(z) } + z P^{R,T}_{gq}(z) \right]  \nonumber \\
 & = &
 \frac{\as}{2\pi}
  \int_x^{1} \frac{dz}{z} \left[ z {P}^{R,T}_{qq} \left(z\right) \frac{x}{z}{D_q}\left({\frac{x}{z}}, \mu^{2}\right) +   {z P}^{R,T}_{gq} \left(z\right) \frac{x}{z}{D_g}\left({\frac{x}{z}}, \mu^{2}\right) \right]  \nonumber \\
  & & -  \frac{\as}{2\pi}  xD_q(x,\mu^2) \int_0^1 dz \, \left[ z{ P}^{R,T}_{qq}(z)  + z P^{R,T}_{gq}(z) \right]  \label{DGLAP_quark_expanded} \, ,
  \end{eqnarray}
where we used in the second line Eq.~\eqref{eq:plus-z-moment} for $P_{qq}$, and all upper
integration limits are understood as $1-\epsilon$.

For the gluon, the same steps with $P_{gg}=R_{gg}(z)+k_g\left[\frac{1}{1-z}\right]_++d_g\,\delta(1-z)$ 
lead to"
\begin{eqnarray}
 \mu^2 \frac{{\partial }{x D_g}(x,\mu^2)}{{\partial } \mu^2}
& = &
 \frac{\as}{2\pi} \left[
  \int_x^{1} \frac{dz}{z} z {P}^{(R,T)}_{gg} \left(z\right) \frac{x}{z}{D_g}\left({\frac{x}{z}}, \mu^{2}\right) + {z P}^{(R,T)}_{qg} \left(z\right) \frac{x}{z}{D_\Sigma}\left({\frac{x}{z}}, \mu^{2}\right) \right] \nonumber \\
  & & -  \frac{\as}{2\pi}  xD_g(x,\mu^2) \left[ \int_0^1 dz \, z  {P}^{(R,T)}_{gg}(z)  + 2 n_f  \,  z P^{(R,T)}_{qg}(z) \right]  \, .
  \label{eq:timelike-pb-evolution-gluon}
  \end{eqnarray}

\end{tolerant}

\bibliographystyle{mybibstyle-new.bst}
\raggedright
\input{pb_updfevolv_note.bbl}

\end{document}

%% file: pb_updfevolv_note.bbl
\providecommand{\href}[2]{#2}\begingroup\raggedright\endgroup